\documentclass[twocolumn,superscriptaddress,prb,showpacs,preprintnumbers,amsmath,amssymb]{revtex4}

\usepackage{graphicx}
\usepackage{dcolumn}
\usepackage{bm}
\usepackage{ulem}
\usepackage{color}
\usepackage{verbatim}
\usepackage{amssymb}

\begin{document}

\title{Thin films MnGe grown on Si(111)}

\author{J. Engelke}
\affiliation{Institut f\"ur Physik der Kondensierten Materie, Technische Universit\"at Braunschweig, Mendelssohnstr. 3, D-38106 Braunschweig, Germany}
\author{D. Menzel}
\affiliation{Institut f\"ur Physik der Kondensierten Materie, Technische Universit\"at Braunschweig, Mendelssohnstr. 3, D-38106 Braunschweig, Germany}
\author{V. A. Dyadkin}
\affiliation{Swiss-Norwegian Beamlines at the European Synchrotron Radiation Facility, Grenoble, 38000, France}

\date{\today}

\begin{abstract}
MnGe has been grown as a thin film on Si(111) substrates by molecular beam epitaxy. A 10\,\AA\, layer of MnSi was used as seedlayer in order to establish the B20 crystal structure. Films of a thickness between 45 and 135\,\AA\, have been prepared and structually characterized by RHEED, AFM and XRD. These techniques give evidence that MnGe forms in the cubic B20 crystal structure as islands exhibiting a very smooth surface. The islands become larger with increasing film thickness. A magnetic characterization reveals that the ordering temperature of MnGe thin films is enhanced compared to bulk material. The properties of the helical magnetic structure obtained from magnetization and magnetoresistivity measurements are compared with films of the related compound MnSi. The much larger Dzyaloshinskii-Moriya interaction in MnGe results in a higher rigidness of the spin helix.

\end{abstract}


\maketitle

\section{\label{sec:Intro}Introduction}
The experimental evidence of the theoretically predicted skyrmions in non-centrosymmetric compounds with Dzyloshinskii-Moriya interaction has intrigued many scientists over the last years.\cite{Muehlbauer,Pfleiderer,roessler} Recently, the preparation of thin films of B20 type MnSi on silicon substrates \cite{wir,karhu2012,wilson} has offered promising prospects with regard to possible applications in future spintronic devices. On the one hand MnSi films offer a variety of interesting magnetic phases and on the other hand they are easy to integrate into devices due to the use of silicon as substrate material being well established in technology. The benefit of thin films compared to bulk material is the existence of the skyrmion phase in an extended region of the magnetic phase diagram due to the uniaxial anisotropy.\cite{butenko,li,wilson} This pioneers new opportunities for data storage devices.

The drawback using MnSi films is the low magnetic ordering temperature, which is considerably below liquid nitrogen temperature. Therefore, it is the aim to find compounds with similar spin order at higher temperatures. A suitable candidate is the B20 compound MnGe (bulk lattice constant of 4.795\,\AA ) with a magnetic ordering temperature $T_{ord}$ of 170\,K.\cite{takizawa} The magnetic ground state of MnGe is a helical spin structure with a helix length between 3\,nm at lowest temperatures and 6\,nm near $T_{ord}$.\cite{kanazawa1} The helix axis is due to magnetic anisotropy pinned along $<$001$>$,\cite{Makarova} but rotates into field direction in an applied field. Recently, a large topological Hall effect exceeding by 40 times that of MnSi was observed in MnGe which was attributed to a skyrmion phase in analogy to MnSi.\cite{kanazawa1}. Further evidence for the existence of skyrmions was given by small angle neutron scattering experiments.\cite{kanazawa2}

Unfortunately, the synthesis of MnGe is considerably laborious, since it forms only under high pressure and temperatures between 600 and 1000\,$^\circ$C.\cite{takizawa} However, molecular beam epitaxy (MBE) allows for thin film growth under strong non-equilibrium conditions. Nevertheless, there has been no successfull attempt to grow MnGe on Ge, since Mn and Ge tends to form Mn$_5$Ge$_3$.\cite{olive,gunnella} The use of Si(111) as substrate offers the opportunity to prepare a seedlayer of MnSi, which realizes the B20 crystal structure for MnGe growth. The lattice constant of MnGe within the (111) plane matches that of Si with a misfit of only 2\,$\%$, thus, compressively strained MnGe films may be grown on Si(111) substrates.

In this paper we show a preparation method for MnGe thin films on Si substrates with the aid of a MnSi seedlayer. The structure and morphology of the films have been investigated by reflection high-energy electron diffraction (RHEED), atomic force microscopy (AFM) and X-ray diffraction (XRD). To determine the physical properties of the samples magnetization and magnetoresistance measurements have been performed.

\section{Film growth and structural characterization}
For the growth of MnGe thin films P-doped Si(111) substrates were used, which possess a resistivity between 1 and 10 $\Omega$cm at room temperature. Prior to film deposition the substrates were heated to 1100\,$^\circ$C under UHV conditions in order to remove the oxide layer and to achieve a clean and flat surface with 7$\times$7-reconstruction, which was verified by in-situ RHEED investigations. The depostion of Mn and Ge directly on the Si(111) surface does not produce B20 MnGe films but results in a Mn$_5$Ge$_3$ layer. In order to establish the B20 crystal structure a 5\,\AA\,Mn layer was deposited onto the Si surface and heated to 300\,$^\circ$C subsequently. This procedure provides for the formation of a thin MnSi seedlayer with a thickness of 10\,\AA . In a second step, MnGe is codeposited by simultanoeus evaporation of Mn and Ge from an effusion cell and an electron beam evaporator, respectively. During film growth with a rate of 0.15\,\AA/s the substrate is held at a temperature of 250\,$^\circ$C. 

\begin{figure}[htbp]
	\centering
		\includegraphics[width=0.35\textwidth]{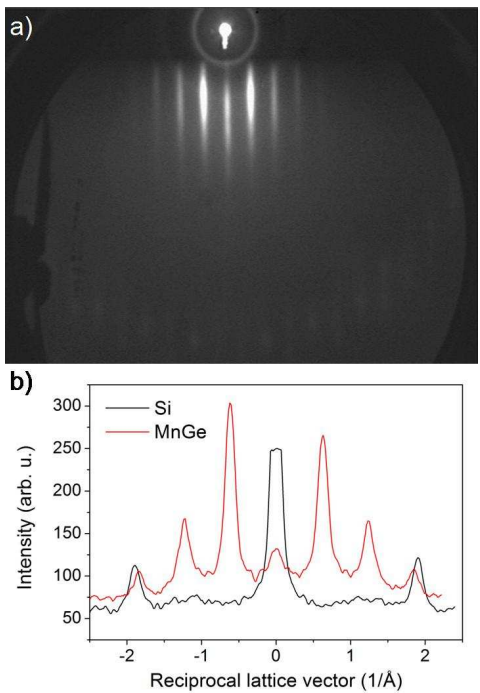}
	\caption{a) RHEED pattern of a 135\,\AA\, film along the $[10\bar1]$ crystal direction and b) line scans across the RHEED streaks for the MnGe film in comparison with the Si substrate. The scans were taken parallel to the shadow edge.}
	\label{fig:RHEED}
\end{figure}

The MnGe films have been investigated by in-situ RHEED in order to determine their structure and morphology. The RHEED pattern of a 135\,\AA\, MnGe film observed along the $[10\bar1]$ direction of the Si substrate indicates two-dimensional film growth [Fig. \ref{fig:RHEED}(a)]. The arrangement of the streaks is very similar to the pattern of MnSi thin films,\cite{wir} and suggests that MnGe sustains the B20 crystal structure provided by MnSi seedlayer. The uniformity of the intensity of the detected streaks implies a flat surface of a size comparable to the area contributing to the RHEED pattern of around 100\,nm in diameter.

Line scans across RHEED patterns of a 135\,\AA\, MnGe film [Fig. \ref{fig:RHEED}(b)] compared to the Si substrate reveal a nearly pseudomorphic growth of the MnGe layer. However, a small deviance of the MnGe streaks from the corresponding Si reflections indicates that the MnGe lattice has at least partly relaxed from the compressive strain imposed by the substrate. \\

\begin{figure}[ht]
	\centering
		\includegraphics[width=0.4\textwidth]{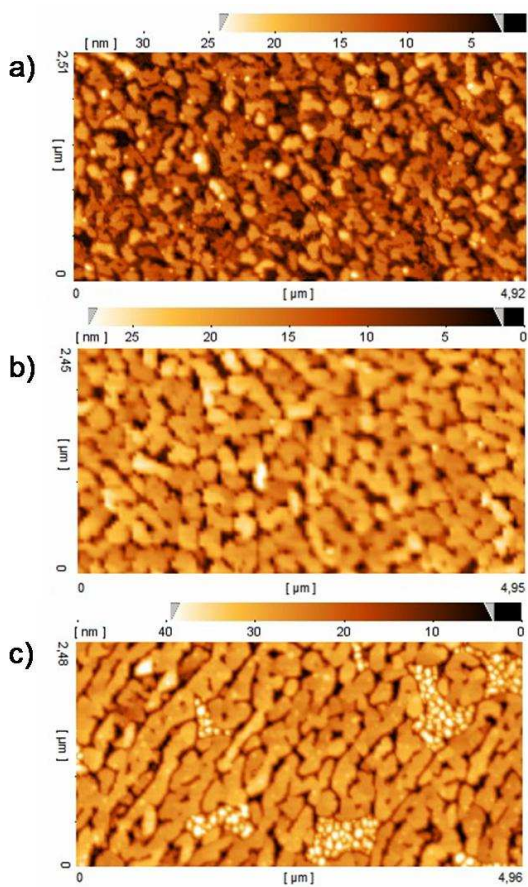}
	\caption{AFM images of MnGe films grown on Si(111) with a thickness of a) 45\,\AA , b) 90\,\AA, and c) 135\,\AA .}
	\label{fig:AFM}
\end{figure}
AFM images of films with thicknesses of 45, 90 and 135\,\AA\, give evidence that island growth of Vollmer-Weber type is the predominant growth mode [Fig. \ref{fig:AFM}]. The thinnest film of 45\,\AA\, thickness [Fig. \ref{fig:AFM}(a)] consists of islands with a typical diameter of 100\,nm separated by valleys of similar size. With increasing film thickness the islands are enlarged and gradually fill the space between them. For the 135\,\AA\, film only very thin valleys of a few nm can be observed [Fig. \ref{fig:AFM}(c)], and the morphology has transformed into elongated islands with a length of up to 2\,$\mu$m and a width of around 200\,nm.
\begin{figure}[htbp]
	\centering
		\includegraphics[width=0.40\textwidth]{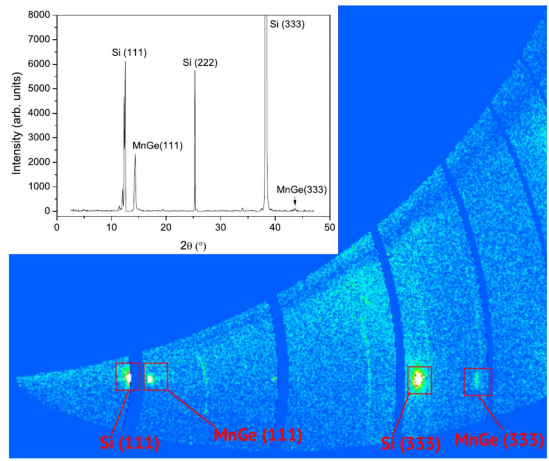}
	\caption{XRD measurement on a 135\,\AA\, sample MnGe. The plane of the picture is spanned by the $[111]$ and $[0\bar1 1]$ crystal directions. Inset: Intensity plot along the $[111]$ direction.}
	\label{fig:xray}
\end{figure}

X-ray diffraction measurements were performed using synchrotron radiation with $\lambda = 0.6824$\,\AA\ at the Swiss-Norwegian Beamline BM1A of the ESRF (Grenoble, France) with the {\sc pilatus@snbl} diffractometer. The investigation of the 135\,\AA\, film confirms the B20 crystal structure of the MnGe. In Fig. \ref{fig:xray} the (111) and (333) peaks of the Si substrate and the MnGe thin film are clearly resolved as single crystal peaks. The inset shows an integrated  diffraction pattern along the [111] direction. From the position of the MnGe(111) peak the lattice parameter of the MnGe film of (4.750 $\pm$ 0.004)\,\AA\ is obtained, which is 1\% smaller than the value for bulk MnGe due to compressive strain. \cite{takizawa}

\section{Magnetic measurements}
\begin{figure}[htbp]
	\centering
		\includegraphics[width=0.5\textwidth]{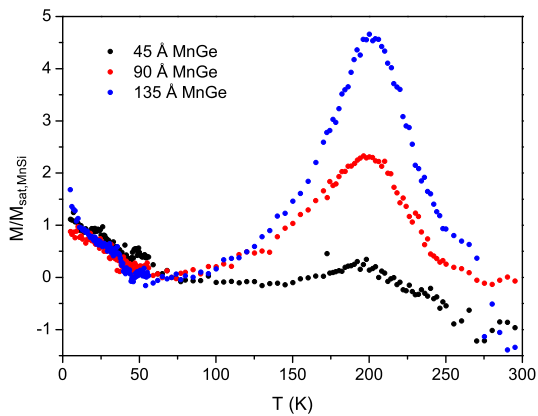}
	\caption{Temperature dependent magnetization of different MnGe films measured in a magnetic field of 10\,mT. The data are normalized with respect to the saturation magnetization of the MnSi seedlayer.}
	\label{fig:susceptibility}
\end{figure}
The magnetic characterization of the MnGe films was carried out using a {\sc Quantum Design} MPMS-5S SQUID magnetometer. For films of different thickness the temperature dependence of the magnetic susceptibility has been measured in the range from 5\,K to 300\,K in an applied magnetic field of 10\,mT [Fig. \ref{fig:susceptibility}]. Below 40\,K the susceptibility slightly increases due to the MnSi seedlayer, that orders magnetically in this temperature range. The measurements were normalized with respect to the saturation magnetization of the MnSi seedlayer, because this layer is the same for all three films. The susceptibility of MnGe films exhibits an ordering temperature of $T_{ord} = 200$\,K  indicated by a broad peak.  Regarding MnGe bulk material, the susceptibility shows a qualitatively similar behavior with a lower $T_{ord} \approx 170\,$K.\cite{kanazawa1} An enhancement of the ordering temperature has also been observed for MnSi thin films.\cite{wir,karhu2010} In contrast to MnSi thin films, no thickness dependence of $T_{ord}$ can be detected for films between 45 and 135\,\AA. Possibly, the spin-spin correlation length is shorter than the value for MnSi films (7~monolayers) \cite{wir}, i.\,e. the thickness dependence may occur for MnGe when the films are thinner than investigated in this work.

Both materials belong to the B20 compounds, which possess a helical spin structure, since their magnetic properties are governed by the interplay of ferromagnetic exchange and Dzyaloshinskii-Moriya interactions. Nevertheless, the susceptibility of MnGe shows a behavior which is typical for antiferromagnetic order, whereas for MnSi an increase to a constant value of magnetization towards low temperature is observed. The helix length in MnSi is very long (18\,nm) and, thus, the local magnetic structure is related to ferromagnetism. A small field easily deforms the soft helix and induces a net magnetization. In the case of MnGe the helix is more rigid. Therefore, at low temperature no net magnetic moment is induced by a field as small as 10\,mT. The helix wavelength is connected to the Dzyaloshinskii constant $D$ via $\lambda = 4\pi A/D$, where $A$ is the magnetic stiffness.\cite{roessler} Since the helix in MnGe is extremely short ($\approx 3$\,nm)\cite{kanazawa1} the Dzyaloshinskii constant is expected to be large, and the magnetic structure is very close to an antiferromagnet.

\begin{figure}[htbp]
	\centering
		\includegraphics[width=0.50\textwidth]{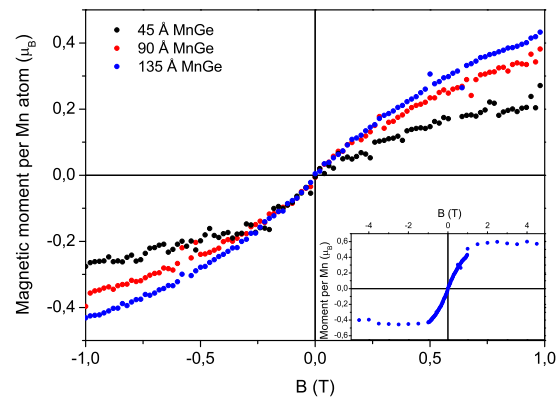}
	\caption{Magnetization curves performed at 200\,K on thin MnGe films of different thickness. Inset: Magnetization at 200\,K up to 5\,T for the 135\,\AA\, film.}
	\label{fig:magnetization}
\end{figure}

Field dependent magnetization measurements at $T = 200$\,K were carried out on the same three samples as in Fig. \ref{fig:susceptibility}. For all samples the magnetization increases in fields up to 1\,T [Fig. \ref{fig:magnetization}]. The inset of Fig. \ref{fig:magnetization} shows a magnetization measurement on the 135\,\AA\, film in fields up to 5\,T, which reveals that saturation is reached around 1\,T. This is in agreement with measurements performed on bulk MnGe at temperatures close to $T_{ord}$.\cite{kanazawa1} Since the helix length is much shorter than the size of the MnGe islands, the magnetization behavior is not expected to be different from the bulk.

The magnetic moment per Mn atom was calculated assuming that the complete amount of Mn deposited during growth has reacted to MnGe. However, since the magnetic moment is only half of the bulk value, some part of the deposited Mn did not form MnGe. Furthermore, we observe an apparently larger magnetic moment for thicker films. This can be explained by the fact that especially in the beginning of MnGe growth not every Mn atom is incorporated into the MnGe crystal. Evidence for this is also given by magnetization measurements at 5\,K, where mainly the ordering of the MnSi seedlayer is observed, since MnGe only gives a linear contribution in the considered field range. We observe a moment of the MnSi layer that is about twice as large as expected for 10\,\AA\, MnSi. Thereby, we conclude that some part of the deposited Mn has reacted with Si from the substrate to form MnSi.\\

\section{Magnetoresistivity}
Resistivity measurements were performed on the 135\,\AA\, film using the van-der-Pauw method. The sample was found to be metallic, and the residual resistivity at 3\,K was determined as 83\, $\mu \Omega$cm. 
\begin{figure}[htbp]
	\centering
		\includegraphics[width=0.50\textwidth]{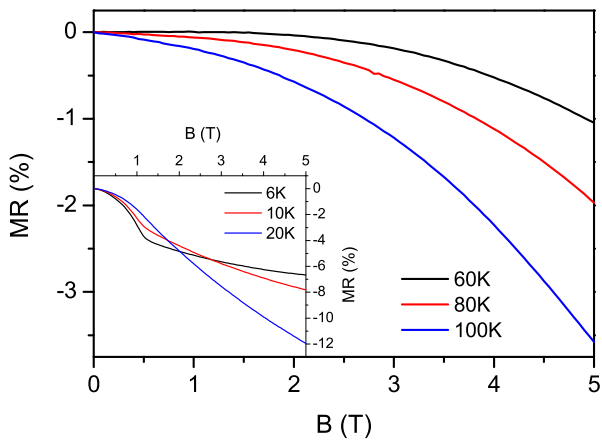}
	\caption{Magnetoresistivity of a 135\,\AA\, MnGe film measured at different temperatures. Inset: Comparison with equivalent measurements on a 19\,nm MnSi film.}
	\label{fig:MR}
\end{figure}
The field dependence of the resistivity was measured in magnetic fields up to 5\,T for several temperatures. In Fig. \ref{fig:MR} three curves are depicted, which represent the magnetoresistivity defined by $MR=100\cdot(\rho(B)-\rho(0))/\rho(0)$. The data were obtained at temperatures between 60\,K and 100\,K, where the sample is in a magnetically ordered state. The MR effect is negative for all temperatures and fields and exhibits no remarkable features in the investigated field range. For comparison the equivalent data for a 19\,nm film MnSi is shown in the inset of Fig. \ref{fig:MR}. In the case of MnSi the critical magnetic field, where the spins align ferromagnetically, occurs around 1\,T. At this field a clear kink accompanied by a change in curvature is observed. Furthermore the size of the MR effect is considerably larger for MnSi. Regarding MnGe the absence of magnetic phase transitions in moderate magnetic fields and the smallness of the magnetoresistivity evidences that the helical structure is more rigid than in MnSi. As discussed in the previous paragraph this can be ascribed to a stronger Dzyaloshinskii-Moriya interaction.

\section{Conclusion}
In this work we have proved that we succeeded in growing crystalline MnGe as a thin film on a Si(111) substrate. The film adopts the B20 structure from a thin seedlayer of MnSi prepared prior to MnGe growth. Morphological investigations using RHEED and AFM give evidence that the MnGe thin films consist of islands with a flat surface, which enlarge during growth. The B20 structure was confirmed by XRD and the lattice parameter was determined to be 1\% smaller than in bulk MnGe due to compressive strain imposed by the Si substrate. \\
Although the magnetic properties of MnGe thin films are found to be qualitatively similar to bulk, the ordering temperature is enhanced to 200\,K. In magnetoresistivity measurements no critical fields were observed up to 5\,T. Compared to MnSi, the helix in MnGe is shorter and more rigid than in MnSi. Therefore, the magnetic structure is related to antiferromagnetism rather than to ferromagnetism. 

\acknowledgments
We would like to thank Dmitry Chernyshov for his support with the X-ray measurements at the European Synchrotron Radiation Facility. The AFM measurements were performed at the Institute of Semiconductor Technology in Braunschweig. We thank Alexander Wagner for his help with the equipment.

\end{document}